\documentclass{moriond}

\def\be{\begin{equation}}
\def\ee{\end{equation}}
\def\bea{\begin{eqnarray}}
\def\eea{\end{eqnarray}}

\begin{document}

\title{Implications of a Dark Matter-Neutrino Coupling at Hyper--Kamiokande}

\author{\textbf{Andr\'es Olivares-Del Campo} \footnote{$\,$Speaker}}

\address{Institute for Particle Physics Phenomenology, Durham University, South Road, Durham, DH1 3LE, United Kingdom}

\author{Sergio Palomares-Ruiz}

\address{Instituto de F\'{\i}sica Corpuscular (IFIC), CSIC-Universitat de Val\`encia, Apartado de Correos 22085, E-46071 Val\`encia, Spain}

\author{ Silvia Pascoli}

\address{Institute for Particle Physics Phenomenology, Durham University, South Road, Durham, DH1 3LE, United Kingdom}

\maketitle\abstracts{
Dark matter and neutrinos provide the two most compelling pieces of evidence for new physics
beyond the Standard Model of Particle Physics but they are often treated as two different sectors. In this paper, we consider how neutrino observables can be used to constrain the parameter space of different models where active neutrinos interact with DM via mediators of different spins. We study for the first time the sensitivity of the Hyper--Kamiokande detector to neutrinos produced from MeV Dark Matter annihilation. In particular, we find that thermally produced DM candidates with masses between 15 -- 30 MeV could be fully excluded.
}

\section{Introduction}
The study of Dark Matter (DM) has been the focus of many articles in the last forty years. Nevertheless, a Particle Physics description of the nature of DM is still missing. It is known, however, that DM must have interactions in order to be produced in the early Universe. An appealing possibility is to consider DM interacting with neutrinos. These interactions can also account for the generation of neutrino masses in some models \cite{Boehm:2006mi}, and lead to observable signatures in the Early Universe and in neutrino detectors. In this paper we first describe the relevant observational signatures of DM-$\nu$ interactions in Section 2, focusing on the prospects of detecting at Hyper--Kamiokande (HK) the neutrinos produced from DM annihilation. This new study significantly improves on previous results obtained using data from the Super--Kamiokande detector. In Section 3, we show an example of two simplified models of Dirac DM interacting with active neutrinos and how the complementarity between different observables is a powerful tool to constrain the parameter space of such models. We conclude in Section 4. 

\section{Experimental Signatures}
\label{signatures}
DM-$\nu$ interactions induce a variety of experimental signatures, either through the elastic scattering between DM and neutrinos or via the annihilation of two DM particles into a neutrino/antineutrino pair. The former has an impact on the cosmic microwave background (CMB) and leads to a suppression of large scale structures (LSS) in our Universe. By confronting CMB and LSS predictions to observations, one can get an upper bound on the strength of the DM-$\nu$ elastic scattering for a temperature-dependent cross section~\cite{Wilkinson:2014ksa} 
\begin{equation}
\sigma_{\rm{el}}< 10^{-48} \ \left(\frac{m_{\rm{DM}}}{\rm{MeV}}\right) \ \left(\frac{E_\nu}{E_0}\right)^2 \  \rm{cm^2} \, \, ,
\end{equation}
where $E_0 \sim 2.35 \times 10^{-4}$~eV is the neutrino energy today.

On the other hand, DM annihilation into neutrinos sets the DM relic abundance in a thermal scenario, it could increase the number of relativistic degrees of freedom during big bang nucleosynthesis\cite{Boehm:2013jpa}, and might lead to a monochromatic neutrino signal which could be observed at neutrino detectors. The later occurs because the neutrino produced from DM annihilations in high density regions carries away an energy equal to the DM mass. 

It has been shown that neutrino experiments with a low--energy threshold such as the SK detector in Japan, can set limits on the annihilation cross section of DM particles with MeV masses by studying the inverse beta decay (IBD) events for MeV neutrino energies \cite{PalomaresRuiz:2007eu,Campo:2017nwh,Primulando:2017kxf}. Here we follow an analogous analysis and consider the expected neutrino signal from DM annihilation in the Milky way at the HK detector. Similarly, we simulate the dominant backgrounds for energies between $10-130$ MeV, which are given by the atmospheric $\nu_e$ (and $\bar\nu_e$) flux and the Michel positrons (and electrons) from the decays at rest of muons with energies below the detection threshold produced by atmospheric neutrinos (i.e., invisible muons). 

HK is a water Cherenkov detector with a fiducial volume of 560 kton which is planned to be built in Japan within the next decade. The expected number of neutrino events at HK for MeV energies are estimated using a scaling in fiducial volume of a factor of $\sim 25$ with respect to the MeV neutrino data recorded during the SK phases I--III. In addition, studies show that adding 0.1\% by mass of gadolinium (Gd) to the water detector would allow to tag the neutron produced in the IBD process which in turn, would reduce the background due to invisible muons by a factor of 5 \cite{Abe:2011ts}. In this analysis, we use for HK the same energy resolution and efficiency as the ones achieved during the different SK phases. The energy resolution is an important experimental parameter when determining the sensitivity of such monochromatic searches and therefore, our limits will be weaker if HK has a lower energy resolution than SK.

\begin{figure}[h!]
\centerline{\includegraphics[width=0.8\linewidth, height=6.4cm]{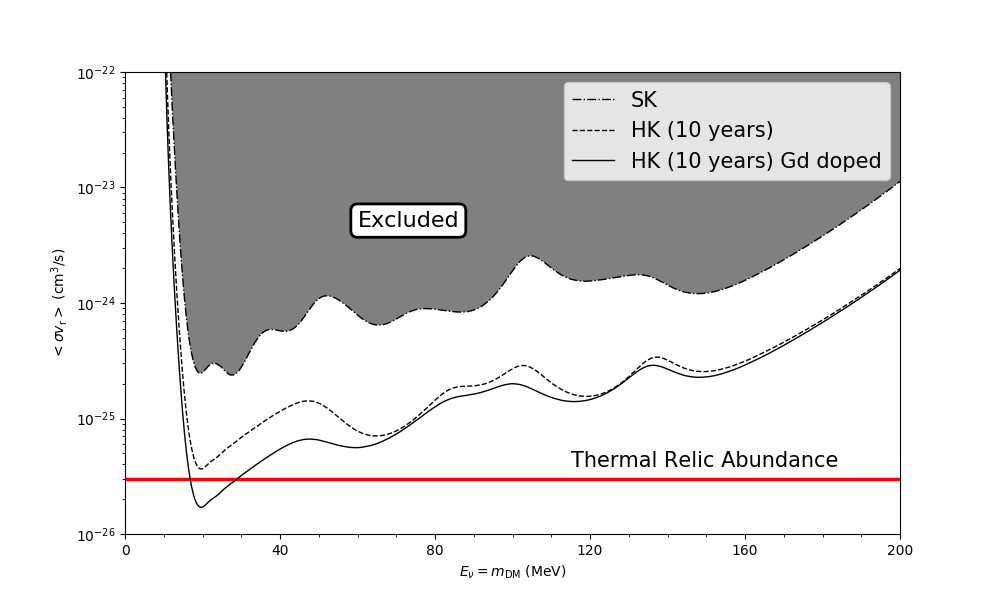}}
\caption[]{90\% limits on the thermally averaged DM annihilation cross section as a function of the DM mass. The solid and dashed line represent the expected annihilation rate at HK for a data taking period of 10 years, where in the former we have considered the effects of Gd doping to the detector, assuming that the invisible muon background is reduced by 80\%. The dot-dashed line are the results from \cite{Campo:2017nwh}. The thick red line corresponds to the value that is needed to explain the observed abundance in thermal DM scenarios, i.e., $\langle \sigma v_r \rangle = 3\times 10^{-26}\; \rm{cm}^3/\rm{s}$. }
\label{fig:HK}
\end{figure}

In the absence of a DM signal, we can compute the 90\% confidence level upper limit on the number of neutrinos produced from DM annihilation during a total time exposure of 10 years. This limit can be converted into a 90\% limit on the thermally averaged DM annihilation cross section to neutrinos (Fig.\ref{fig:HK}). This figure shows that HK will be able to probe DM annihilation cross sections one order of magnitude larger than the analysis done using SK data. Moreover, including Gd would allow to discover or rule out models where DM annihilates into neutrinos for DM masses between $15-30$ MeV.

\section{Dirac DM Models}
We will now demonstrate how these experimental constraints can be used to probe the parameter space of a sub-set of DM models. Following a simplified model approach, we consider the interactions between a DM candidate and active neutrinos via a particular mediator. There is a total of twelve different combinations of spins of the DM and mediator particles consistent with Lorentz invariance. Here, we only study the results for a Dirac DM candidate coupling to neutrinos via scalar or a vector mediator and refer the interested reader to \cite{Campo:2017nwh}. The relevant terms in the Lagrangian are
\begin{eqnarray}
\mathcal{L}_{\rm{int}} &\supset &-\,g\,\phi \, \overline{\chi_{\rm R}} \, \nu_{\rm{L}}    \ +\ \rm{h.c.}  \,\,\,\,\,\,\,\,\,\,\,\,\,\,\,\,\,\,\,\,\,\,\,\,\,\,\,\,\,\,\,\,\,\,\,\,\,\,\,\,\,\,\,\,\,\,\,\,\,\,\,\,\,\,\,\,\,\,\,\,\,\,\,\,\,\,\,\rm{Scalar~ mediator}, \nonumber \\
\mathcal{L}_{\rm{int}}&\supset &- \,g_\nu\overline{\nu_{\rm L}}\gamma^\mu Z'_\mu \nu_{\rm{L}}\, -  g_{\chi} \overline{\chi}\gamma^\mu Z'_\mu \chi \,\,\,\,\,\,\,\,\,\,\,\,\,\,\,\,\,\,\,\,\,\,\,\,\,\,\,\,\,\,\,\,\,\,\,\,\,\,\,\,\,\,\, \rm{Vector~ mediator}.
\end{eqnarray}  

The cross sections are summarized in the table below
\renewcommand{\arraystretch}{2.4}
\begin{table}[h!]
\centering
\begin{tabular}{  c |  c  c c  }
 & \hspace{0.2cm} \textbf{Scalar Mediator} & \hspace{0.2cm} \textbf{Vector Mediator} \\
\hline 
$<\sigma v_{\rm{r}}> \, \, \propto$ & $\frac{g^4m^2_{\rm{DM}}}{(m^2_{\rm{DM}}\, +\, m^2_{\phi})^2}$ & $\frac{ g_\chi^2g_\nu^2 m^2_{\rm{DM}}}{(4m^2_{\rm{DM}} \, -\,m^2_{\rm{Z'}})^4}$  \\ 
$~~\sigma_{\rm{el}} \,\, \propto$ & $  \frac{g^4E^2_\nu}{(m^2_{\rm{DM}}-m^2_{\phi})^2}$ & $  \frac{ g_\chi^2g_\nu^2 E^2_\nu}{m^4_{\rm{Z'}}}$  \\  
\hline 
\end{tabular}
\caption{Relevant terms in the elastic scattering and thermally averaged annihilation cross section expressions for a Dirac DM candidate coupled to neutrinos via a scalar or a vector mediator. $E_\nu$ is the neutrino energy today. }
\label{Table1}
\end{table}

\begin{figure}[h!]
\begin{minipage}{0.50\linewidth}
\centerline{\includegraphics[width=1.1\linewidth, height=6.8cm]{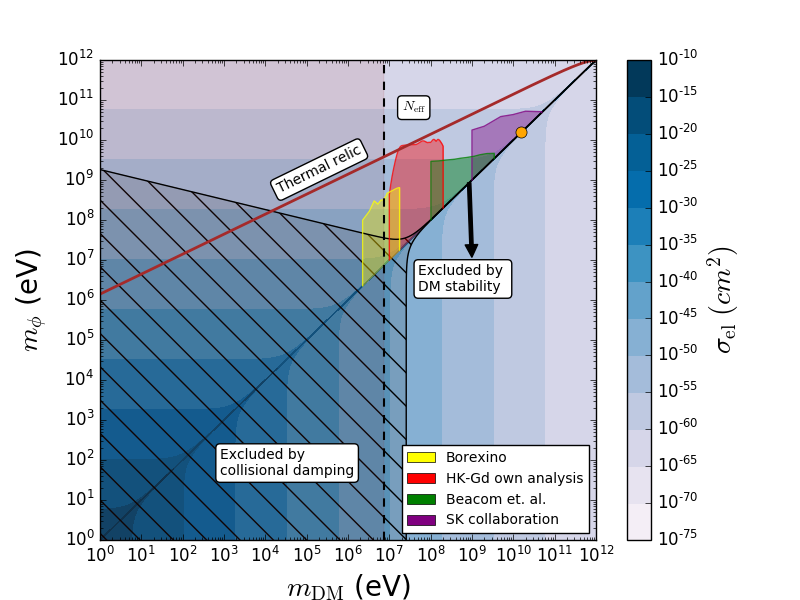} }
\end{minipage}
\hfill
\begin{minipage}{0.50\linewidth}
\centerline{\includegraphics[width=1.1\linewidth, height=6.8cm]{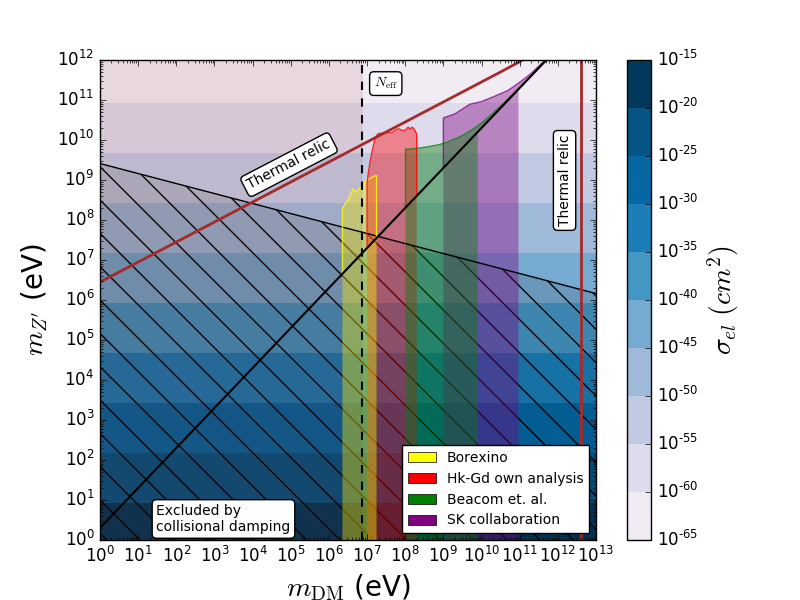} }
\end{minipage}
\caption[]{Elastic scattering of Dirac DM coupled to a scalar (left) and to a vector (right) mediator in the $m_{\rm{Mediator}}-m_{\rm{DM}}$ plane for $g=1$. Different regions are constrained by: the collisional damping limit (dashed region and black line along the diagonal up to the orange dot), a bound from the antineutrino flux at Borexino~\cite{Bellini:2010gn} (in yellow), our analysis at HK with Gd doping described in Section~\ref{signatures} (in red), the analysis done in Ref.~\cite{Yuksel:2007ac} using results from SK, Fr\'ejus and Amanda (in green), and the analysis done by the SK collaboration for GeV neutrinos produced at the galactic centre~\cite{Frankiewicz:2015zma} (in purple). The parameters that give rise to the right relic abundance (brown line) are shown as a reference. The dashed line refers to the upper bound on the DM mass derived from $N_{\rm{eff}}$~\cite{Boehm:2013jpa} \label{Fig2Paper2}} 
\end{figure}

We see that the phenomenology between these two scenarios is very different. When we consider a scalar mediator, $m_{\rm{Mediator}}>m_{\rm{DM}}$ is required to guarantee the stability of the DM particle. This implies that half of the parameter space in the $m_{\rm{Mediator}}-m_{\rm{DM}}$ plane is excluded (see left of Fig. \ref{Fig2Paper2}). The elastic scattering can become resonantly enhanced if the mediator and DM are nearly degenerate and, as can be seen in Ref. \cite{Wilkinson:2014ksa}, it is energy-independent, leading to the exclusion of DM and mediator masses along the diagonal up to $\sim 20$ GeV (i.e. the orange dot on the left of Fig. \ref{Fig2Paper2}).
On the other hand, if the mediator is a spin-1 particle, it can be lighter than the DM (see right of Fig. \ref{Fig2Paper2}), and the annihilation cross section can become resonant.  In this case, the elastic scattering cross section is independent of the DM mass. 

\section{Conclusion}
In this paper, we have studied the prospects of detecting a monochromatic neutrino flux from DM annihilation at HK and we have found that such an indirect detection search could rule out models with thermal DM masses between $15-30$ MeV. Furthermore, we have focused on two Dirac DM models coupled to a scalar or a vector mediator and constrained their $m_{\rm{Mediator}}-m_{\rm{DM}}$ parameter space by imposing the stability of the DM candidate and also, that the DM-$\nu$ interactions are compatible with LSS formation and do not significantly change the CMB angular power spectrum or lead to observable signals at neutrino detectors. We have found that their phenomenology is significantly different and that the new HK constraints play an important role in determining the viability of such models.  We note that for other models in which the annihilation cross section is velocity dependent the constraints derived from HK are considerably weaker due to the small velocity of DM particles in the halo today.

\section*{Acknowledgments}
AO and SP are supported by the  European  Research  Council  under  ERC  Grant ``NuMass''  (FP7-IDEAS-ERC  ERC-CG  617143). SPR is supported by a Ram\'on y Cajal contract, by the Spanish MINECO under grants FPA2014-54459-P, FPA2017-84543-P and SEV-2014-0398, and by the Generalitat Valenciana under grant PROMETEOII/2014/049. SP acknowledges partial support from the Wolfson Foundation and the Royal Society. SP and SPR are also partially supported by the European Union's Horizon 2020 research and innovation program under the Marie Sk\l odowska-Curie grant agreements No. 690575 (RISE InvisiblesPlus) and 674896 (ITN Invisibles). SPR is also partially supported by the Portuguese FCT through the CFTP-FCT Unit 777 (PEst-OE/FIS/UI0777/2013).

\section*{References}

\bibliography{Ref.bib}

\end{document}